\documentclass[journal=nalefd,manuscript=letter]{achemso}

\usepackage[active]{srcltx}
\usepackage{graphicx}
\usepackage{mathcomp}	
\usepackage{hyperref}
\usepackage{color}
\usepackage{amsmath,amssymb,amstext,graphicx,bm}

\author{N.~Ares}
\affiliation
{SPSMS/LaTEQS, CEA-INAC/UJF-Grenoble 1, 17 Rue des Martyrs, 38054 Grenoble Cedex 9, France}

\author{G.~Katsaros}
\affiliation
{SPSMS/LaTEQS, CEA-INAC/UJF-Grenoble 1, 17 Rue des Martyrs, 38054 Grenoble Cedex 9, France}
\alsoaffiliation
{Institute for Integrative Nanosciences, IFW Dresden, Helmholtzstr.\ 20, D-01069 Dresden, Germany}
\alsoaffiliation
{Johannes Kepler University, Institute of Semiconductor and Solid State Physics, Altenbergerstr. 69, 4040 Linz, Austria}
\email{georgios.katsaros@jku.at}

\author{V.~N.~Golovach}
\affiliation
{Institute for Integrative Nanosciences, IFW Dresden, Helmholtzstr.\ 20, D-01069 Dresden, Germany}
\alsoaffiliation
{Centro de F\'{i}sica de Materiales CFM/MPC (CSIC-UPV/EHU) and Donostia International Physics Center DIPC, E-20018 San Sebasti\'{a}n, Spain}
\alsoaffiliation
{IKERBASQUE, Basque Foundation for Science, E-48011 Bilbao, Spain}

\author{J.~J.~Zhang}
\affiliation
{Institute for Integrative Nanosciences, IFW Dresden, Helmholtzstr.\ 20, D-01069 Dresden, Germany}

\author{A.~Prager}
\affiliation
{SPSMS/LaTEQS, CEA-INAC/UJF-Grenoble 1, 17 Rue des Martyrs, 38054 Grenoble Cedex 9, France}

\author{L.~I.~Glazman}
\affiliation
{Department of Physics, Yale University, New Haven, Connecticut 06520, USA}

\author{O.~G.~Schmidt}
\affiliation
{Institute for Integrative Nanosciences, IFW Dresden, Helmholtzstr.\ 20, D-01069 Dresden, Germany}
\alsoaffiliation
{Center for Advancing Electronics Dresden, TU Dresden, Germany}

\author{S.~De~Franceschi}
\affiliation
{SPSMS/LaTEQS, CEA-INAC/UJF-Grenoble 1, 17 Rue des Martyrs, 38054 Grenoble Cedex 9, France}
\email{silvano.defranceschi@cea.fr}

\title{SiGe quantum dots for fast hole spin Rabi oscillations}

\begin{document}

\begin{abstract}
We report on hole g-factor measurements in three terminal SiGe self-assembled quantum dot devices with a top gate electrode positioned very close to the nanostructure. 
Measurements of both the perpendicular as well as the parallel g-factor reveal significant changes for a small modulation of the top gate voltage. 
From the observed modulations we estimate that, for realistic experimental conditions, hole spins can be electrically manipulated with Rabi frequencies in the order of $100$MHz. 
This work emphasises the potential of hole-based nano-devices for efficient spin manipulation by means of the g-tensor modulation technique.
\end{abstract}



Important progress has been made in the past years in the coherent manipulation of confined spins in semiconductor quantum dots (QDs) 
by means of oscillating magnetic and electric fields\cite{Hanson2007,Zwanenburg2013}. 
Spin states can be electrically manipulated either by electric-dipole spin resonance (EDSR)~\cite{Golovach2006,Nowack2007,Nadj-Perge2010} 
or by the so-called g-tensor modulation technique~\cite{Kato2003,Salis2001}.

The EDSR technique is based on the fact that the ac electric field shifts the orbital wavefunction back and forth; 
in an external static magnetic field, due to the presence of a strong spin-orbit coupling, this oscillatory motion induces coherent spin rotations. 
On the other hand, in the g-tensor modulation technique, spin rotations result from an electrically induced oscillation of the Zeeman vector. 
Electrically tunable g-factors are thus essential for this technique.
 
In the past decade, g-factors in different QD systems have been studied thoroughly~\cite{Deacon2011,Csonka2008,Houel2013,Roddaro2007,Nilsson2009,Katsaros2010,Ares2013}.
Recently, Deacon et al.~\cite{Deacon2011} reported electrically tunable electron g-factors for self-assembled InAs nanocrystals (NCs) and estimated the electron Rabi frequency for their double-gate geometry. Since the gates were positioned rather far from the NCs, however, the resulting Rabi frequencies were estimated to be only around $2$MHz.

Interestingly, holes had not been considered as potential qubits for a long time, 
since their spin relaxation times were expected to be very short due to the strong spin-orbit (SO) interaction. 
In spite of the presence of hyperfine interaction and SO coupling, however, 
experiments with InGaAs QDs have shown spin relaxation times, $T_{1}$, as high as several hundreds of microseconds~\cite{Heiss2007}. 
Since in the absence of hyperfine interaction~\cite{Golovach2004} the spin decoherence time, $T_2$, is predicted to be equal to $2T_1$, 
hole-confinement QDs based on isotopically purified SiGe nanostructures are promising candidates for spin qubits with long coherence time.

Hole-confinement QDs have been realized in Ge/Si core/shell nanowires~\cite{Lu2005,Hu2007}, 
where hole spin relaxation times in the order of $1$ms were recently reported~\cite{Hu2011}. 
In order to realize hole-based spin qubits in these systems, the development of all-electrical efficient techniques for fast spin rotations is essential \cite{Kloeffel2013},
because time-dependent magnetic fields are inefficient at inducing Rabi oscillations for holes~\cite{Sleiter2006}.


In this work we have studied holes confined in SiGe NCs. We have placed a top gate $6-7\,{\rm nm}$ away from the NCs, which allows us to electrically tune the g-factor of the hole state. Placing a top gate so close to the NC has several advantages. Firstly, as the coupling to the QD is very strong, driving of the spin states with Rabi frequencies of the order of $100$MHz can be achieved. In addition, as the gate can be very narrow, its action is local and the dissipation of the high-frequency power is minimized.

\begin{figure}[t]
\includegraphics[width=8.5cm,keepaspectratio]{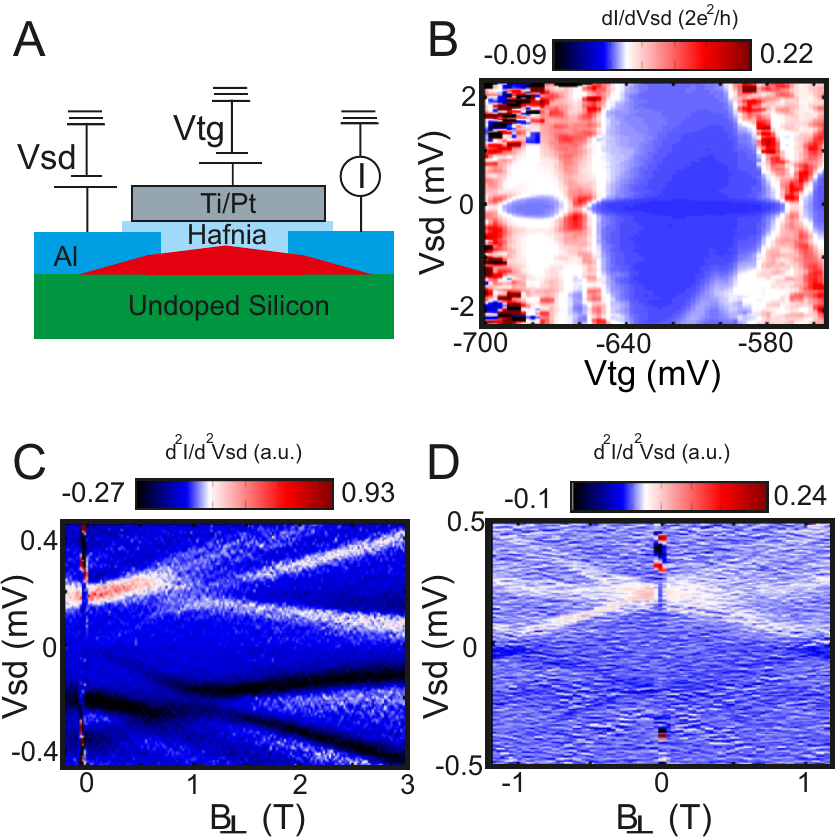}
\caption{ (A) Schematic of a SiGe self-assembled QD device. (B) $dI_{sd}/dV_{sd}$ vs $V_{tg}$ and $V_{sd}$ for
$B=70mT$. The magnetic field is needed for suppressing the superconductivity of the Al electrodes. Inelastic CT steps originating from different orbital levels are present in both diamonds.
(C)-(D) numerical
derivative $d^2I_{sd}/d^2V_{sd}$ vs.
$V_{sd}$ and $B_{\perp}$ for the left and right diamond, respectively. In (C) transitions between two splitted Kramer levels are observed, indicating thus an odd number of localized holes. Transitions between singlet and triplet states are present in (D), implying an even number of confined holes.\cite{Katsaros2010}.}
\label{Fig1}
\end{figure}

The SiGe self-assembled NCs used in this study were grown on undoped (n-) Si(001) wafers.  
The NCs were contacted by aluminium leads [see Fig.\ref{Fig1}(A)] and the top gates were created by depositing $6nm$ of hafnia on top of the obtained devices, followed by deposition of a Ti/Pt $10/90$nm metal electrodes. Low-temperature transport measurements were carried out in a dilution refrigerator with a base temperature of $15\,{\rm mK}$ equipped with accurately filtered wiring and low-noise electronics. A typical differential conductance ($dI_{\it sd}/dV_{\it sd}$) measurement as a function of the top-gate ($V_{\it tg}$) and source-drain ($V_{\it sd}$) voltages is shown in Fig.~\ref{Fig1} (B). A small magnetic field, $B=70\,{\rm mT}$, is applied in order to suppress the superconductivity of the Al electrodes.

Diamond-shaped regions with a charging energy of about $2\,{\rm meV}$ can be observed in Fig.~\ref{Fig1} (B). In the Coulomb blockade regime, single-hole transport is suppressed and electrical conduction is due to second-order cotunneling (CT) processes~\cite{DeFranceschi2001}.
Each CT process involves a hole tunneling out of the QD into the right contact and, simultaneously, another hole entering the QD from the left contact.  At small $V_{sd}$, CT is said to be elastic since it cannot create any excitation in the QD. At sufficiently large $V_{sd}$, however, CT processes can leave the QD in an excited state. Such processes are thus referred to as inelastic. Both diamonds shown here show such inelastic CT features, visible as steps in the $dI_{\it sd}/dV_{\it sd}$. From the position of the inelastic CT step we conclude that the orbital level separation for this device is about $200\,{\rm \mu eV}$.

In order to determine the even or odd occupation of each diamond we performed CT spectroscopy measurements as a function of an applied magnetic field, $B$. These measurements are shown in Fig.~\ref{Fig1} (C)-(D), for the left and right diamond, respectively. A clearly different behavior is observed, revealing the distinct character of the ground states in the two adjacent diamonds. These different behavior of the CT steps as a function of $B$, already reported in ~\cite{Katsaros2010}, is attributed to a left (right) diamond correspondent to an odd (even) number of charges.

\begin{figure}[t]
\includegraphics[width=8.5cm,keepaspectratio]{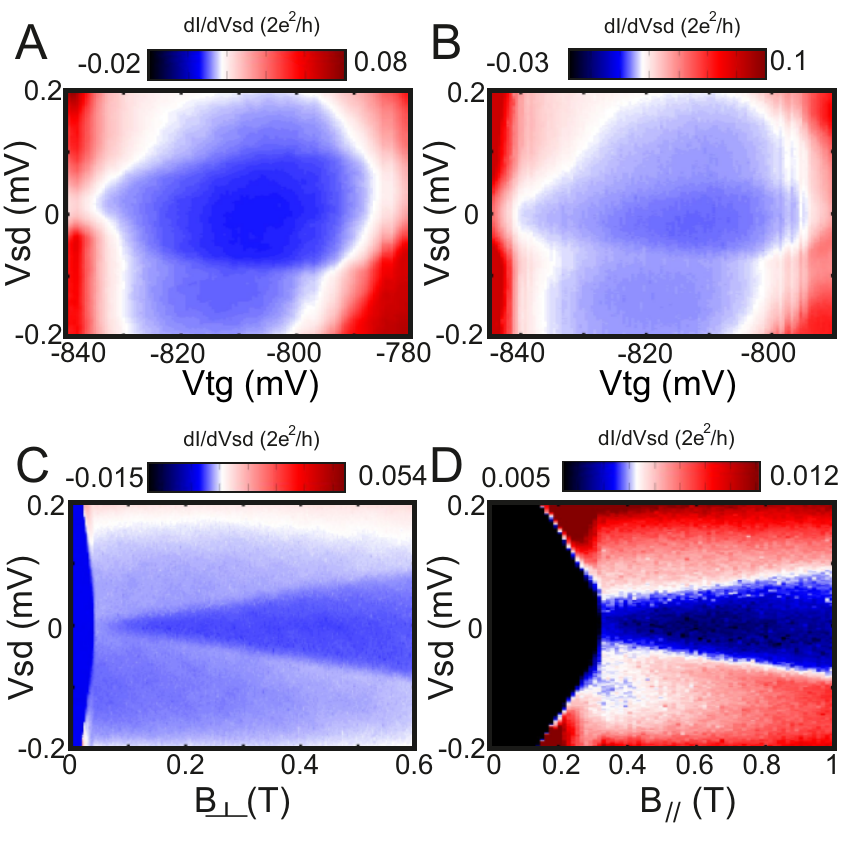}
\caption{(A)-(B) $dI_{sd}/dV_{sd}$ vs. $V_{tg}$ and $V_{sd}$ for $B=0.6T$, applied perpendicular and parallel to the substrate plane, respectively.
(C)-(D) $dI_{sd}/dV_{sd}$ vs. $B$ and $V_{sd}$ for perpendicular and parallel magnetic fields, respectively, demonstrating that the inelastic CT steps are due to the Zeeman splitting of a spin $\frac{1}{2}$ ground state. From the measured Zeeman energies we estimate $g_\perp=(2.0\pm0.2)$ and $g_\parallel=(1.2\pm0.2)$.}
\label{Fig2}
\end{figure}

The differential conductance of the odd diamond was studied further in order to gain more insight into the Zeeman-split Kramers. Figs. ~\ref{Fig2} (A) and ~\ref{Fig2} (B) show a stability diagram of this diamond for $B=0.6$ T applied perpendicular and parallel to the growth plane, respectively. Steps due to the presence of inelastic CT processes are again observed. 

By fixing $V_{tg}$ within the Coulomb blockade regime and sweeping the magnetic field, the behavior of the CT steps was investigated. These measurements are shown in Figs.~\ref{Fig2} (C)-(D). As the CT steps merge together when $B$ approaches zero, we can confirm that the observed steps correspond to the Zeeman splitting of the ground state ($E_Z$). The $g$-factor value perpendicular ($g_\perp$) and parallel ($g_\parallel$) to the substrate plane can be extracted from these measurements since $E_Z=g\mu_BB$, with $\mu_B$ being the Bohr magneton. The extracted values are $g_\perp=(2.0\pm0.2)$ and $g_\parallel=(1.2\pm0.2)$.

Let us now remark on the fact that in both diamonds in Figs.~\ref{Fig2} (A)-(B), the inelastic CT steps are not parallel to the $V_{tg}$ axis. This slope in the CT steps demonstrates that both $g$-factors values are voltage and thus electric-field dependent. From the reported measurements we extract $\frac{\partial g_\parallel}{\partial V_{\rm tg}}=(0.008\pm0.001)\frac{1}{mV}$ and $\frac{\partial g_\perp}{\partial V_{\rm tg}}=(0.007\pm0.001)\frac{1}{mV}$. 


In order to estimate the Rabi frequency, we consider an oscillating voltage $V_{\rm ac}$ superimposed to a constant value $V_{tg}$. Provided that $V_{\rm ac}$ is sufficiently small, the dependence of $g_\parallel$ and $g_\perp$ on $V_{\rm tg}$ can be assumed to be linear and the Rabi frequency of the induced spin rotations reads (see Supporting Information)        

\begin{eqnarray}
f_R &=& \frac{\mu_B V_{\rm ac}}{2h}
\left[
\frac{1}{g_\parallel}\left(\frac{\partial g_\parallel}{\partial V_{\rm tg}}\right)-
\frac{1}{g_\perp}\left(\frac{\partial g_\perp}{\partial V_{\rm tg}}\right)
\right]\nonumber\\
&&\times\frac{g_\parallel g_\perp B_\parallel B_\perp}{\sqrt{\left(g_\parallel B_\parallel\right)^2+\left(g_\perp B_\perp\right)^2}},
\end{eqnarray}
where $h$ is the Planck constant. In this expression, $B_\parallel=B\cos\theta$ and $B_\perp=B\sin\theta$, where the angle $\theta$ is measured with respect to the growth plane. Further, it can be shown that $f_R$ is maximal if $\theta$ is chosen such that
\begin{equation}
\theta_{max} =\arctan\sqrt{\frac{g_\parallel}{g_\perp}}.
\end{equation}

\begin{figure}[t]
\includegraphics[width=8.5cm,keepaspectratio]{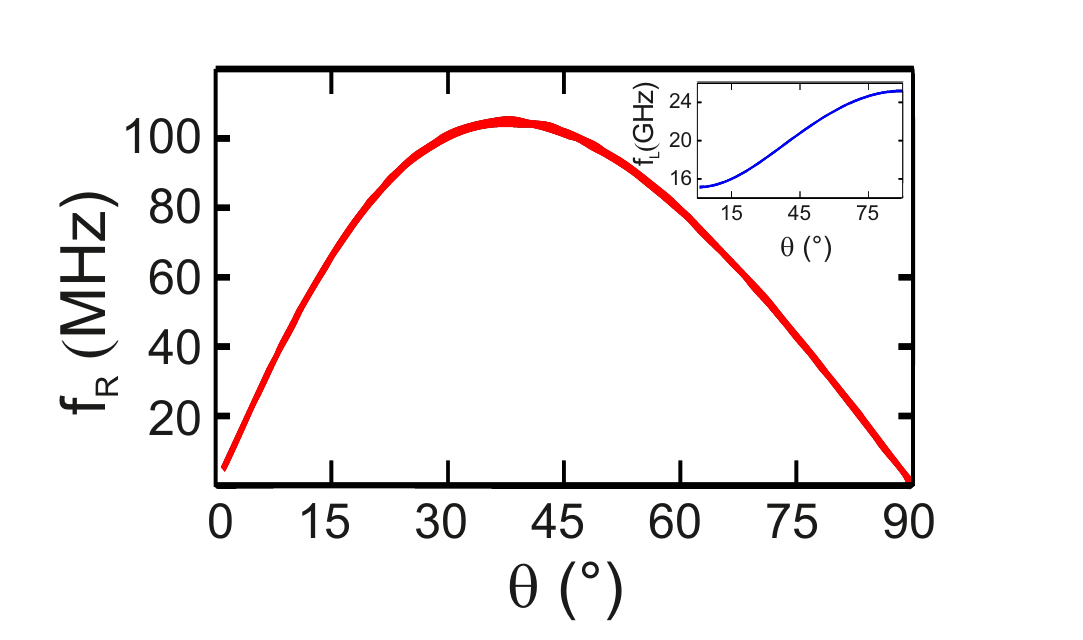}
\caption{Rabi frequency dependence on the magnetic field angle with respect to the substrate plane ($\theta$). It reaches $\sim100MHz$ at $\theta_{max}\approx38^\circ$. The estimated value corresponds to $B=0.9T$ and $V_{ac}=7mV$. Inset: Larmor frequency as a function of $\theta$ for the same experimental conditions. A driving frequency of $\sim20GHz$ is estimated for $B=0.9T$.}
\label{Fig3}
\end{figure}

The experimental values obtained for $g_\parallel$, $g_\perp$, $\frac{\partial g_{\parallel}}{\partial V_{\rm tg}}$ and $\frac{\partial g_\perp}{\partial V_{\rm tg}}$, were used to estimate $f_R$ for a given value of the Larmor frequency ($f_L$) and for different values of $\theta$ (see Fig.~\ref{Fig3}).

By considering a driving frequency $f_L\approx20$ GHz, which corresponds to $B\approx0.9$T applied at an angle $\theta_{max}$, we obtain a Rabi frequency $f_R \sim 100$ MHz for $V_{\rm ac}\approx7$mV. 
Taking the lever-arm parameter of the gate electrode \cite{leverarm}, $\alpha\approx0.07$, this value of $V_{\rm ac}$ corresponds to an energy shift of $\approx500\mu{eV}$, which is of the same order of typically energy-level modulations in EDSR experiments \cite{vandenBerg2013, Nadj-Perge2010}. We note that the estimated value for the Rabi frequency is comparable to values recently reported for electrons confined in InSb nanowires \cite{vandenBerg2013}. 
 
In summary, we have demonstrated that a single top-gate electrode, defined close to a SiGe self-assembled QD, may enable us to perform spin manipulations by means of the g-tensor modulation technique. Our measurements demonstrate that fast Rabi frequencies can be achieved for realistic experimental conditions. The obtained values together with the expectedly long spin coherence times for carriers in Ge underline the potential of holes confined in SiGe QDs as fast spin qubits.

\suppinfo

In order to derive the expression for the Rabi frequency in the main text, 
let us consider a spin $1/2$ driven by the g-tensor modulation.
The time evolution of the spin is governed by the Bloch equation,
\begin{equation}
\frac{d}{dt}\left\langle\bm{S}\right\rangle=
\left[\bm{\omega}_{\rm L}+\delta\bm{\omega}(t)\right]
\times
\left\langle\bm{S}\right\rangle,
\label{eqBloch}
\end{equation}
where $\left\langle\bm{S}\right\rangle$
is the expectation value of the spin 
$\bm{S}=\left(1/2\right)\bm{\sigma}$,
with $\bm{\sigma}$ being the Pauli matrices,
$\bm{\omega}_{\rm L}$ is the Larmor frequency,
and 
$\delta\bm{\omega}(t)$ is the time-dependent frequency part due to the driving field.
In the g-tensor modulation, the driving field arises because of the 
{\it ac} signal sent to the top gate,
\begin{equation}
V_{\rm tg}(t) = V_{\rm tg}^0 + V_{\rm ac}\sin\left(\omega_{\rm ac}t\right),
\end{equation}
where $V_{\rm tg}^0$ is the average value of the top-gate voltage,
$V_{\rm ac}$ is the resulting amplitude of voltage oscillations,
and $\omega_{\rm ac}$ is the {\em ac} angular frequency.
Since we use dome-like SiGe nanocrystals,
which roughly obey rotational symmetry about the axis $z\equiv [001]$,
the g-tensor $\hat{g}$ is approximately diagonal in the main crystallographic frame $(x,y,z)$.
The non-zero elements of $\hat{g}$ include two equal-to-each-other inplane components, $g_x=g_y\equiv g_\parallel$, 
and one out-of-plane component, $g_z\equiv g_\perp$.
The g-tensor modulation can, therefore, be written as
\begin{eqnarray}
g_\parallel(t) &\approx& g_\parallel^0 + \alpha_\parallel V_{\rm ac}\sin\left(\omega_{\rm ac}t\right),\nonumber\\
g_\perp(t) &\approx& g_\perp^0 + \alpha_\perp V_{\rm ac}\sin\left(\omega_{\rm ac}t\right),
\end{eqnarray}
where $g_\parallel^0\equiv g_\parallel$ and $g_\perp^0\equiv g_\perp$ are constant, 
$\alpha_\parallel = \frac{\partial g_{\parallel}}{\partial V_{\rm tg}}$,
and $\alpha_\perp=\frac{\partial g_{\perp}}{\partial V_{\rm tg}}$.
Here, it was assumed 
that $V_{\rm ac}$ is sufficiently small, so that $g_\parallel$ and $g_\perp$ depend linearly on $V_{\rm tg}$ in the voltage window $V_{\rm tg}^0 \pm V_{\rm ac}$.
Thus, the Larmor frequency entering in Eq.~(\ref{eqBloch}) is identified as
\begin{equation}
\bm{\omega}_{\rm L}=
\frac{\mu_{\rm B}}{\hbar}\hat{g}\cdot\bm{B},
\label{eqLarmor}
\end{equation}
whereas the contribution due to driving as
\begin{equation}
\delta\bm{\omega}(t)=
\frac{\mu_{\rm B}}{\hbar}\left(\hat{\alpha}\cdot\bm{B}\right)V_{\it ac}\sin\left(\omega_{\rm ac}t\right).
\label{eqdomgoft}
\end{equation}
To be concise here, we used tensor-vector multiplication, like $\left(\hat{g}\cdot\bm{B}\right)_i=\sum_jg_{ij}B_j$.
The tensor of linear coefficients, $\alpha_{ij}=\partial g_{ij}/\partial V_{\rm tg}$, needs not, in general, be proportional to $\hat{g}$.
Therefore, the time-dependent driving $\delta\bm{\omega}(t)$ may have a component that is transverse to the vector $\bm{\omega}_{\rm L}$,
{\em cf.} Eqs.~(\ref{eqLarmor}) and~(\ref{eqdomgoft}).
This circumstance is at the heart of the g-tensor modulation technique used to induce Rabi oscillations.

It is easiest to solve Eq.~(\ref{eqBloch}) in a frame rotating at frequency $\omega_{\it ac}$ about the vector $\bm{\omega}_{\rm L}$.
The time evolution of the spin is approximated as follows
\begin{eqnarray}
\left\langle S_\pm(t)\right\rangle &\approx&\tilde{S}_\pm(t)e^{\pm i\omega_{\it ac}t},\nonumber\\
\left\langle S_Z(t)\right\rangle &\approx&\tilde{S}_Z(t),
\end{eqnarray}
where $S_\pm=S_X\pm iS_Y$ and the coordinate frame $(X,Y,Z)$ has $Z\parallel \bm{\omega}_{\rm L}$.
The new unknown functions $\tilde{\bm{S}}(t)$ obey a time-independent Bloch equation
\begin{equation}
\frac{d}{dt}\tilde{\bm{S}}=
\left(\bm{\delta}+\bm{\omega}_{\rm R}\right)
\times
\tilde{\bm{S}},
\label{eqBlochtildeS}
\end{equation}
where $\bm{\delta}=\bm{\omega}_{\rm L}\left(1-\omega_{\it ac}/\omega_{\rm L}\right)$ is the detuning from resonance 
and $\bm{\omega}_{\rm R}$ is the Rabi frequency given by~\cite{Golovach2006}
\begin{equation}
\bm{\omega}_{\rm R} =\frac{\mu_B}{2\hbar}V_{\it ac}\left[\left(\hat{\alpha}\cdot\bm{B}\right)\times\bm{n}\right],
\end{equation}
where $\bm{n}=\bm{\omega}_{\rm L}/\omega_{\rm L}$ is the unit vector along the Larmor frequency.

Next, we consider the situation realized in the experiment.
The magnetic field can be rotated in a plane perpendicular to the substrate.
Let us assume that it is the $(y,z)$-plane and represent the magnetic field as
\begin{equation}
\bm{B}={\bm{e}_y}{B_\parallel} + {\bm{e}_z}{B_\perp},
\end{equation}
where $\bm{e}_i$ ($i=x,y,z$) are unit vectors.
Then, the Rabi frequency reads $\bm{\omega}_{\rm R} = {\bm{e}_x}\omega_{\rm R}$, with
\begin{eqnarray}
\omega_{\rm R} &=& \frac{\mu_B V_{\rm ac}}{2\hbar}
\left[
\frac{1}{g_\parallel}\left(\frac{\partial g_\parallel}{\partial V_{\rm tg}}\right)-
\frac{1}{g_\perp}\left(\frac{\partial g_\perp}{\partial V_{\rm tg}}\right)
\right]\nonumber\\
&&\times\frac{g_\parallel g_\perp B_\parallel B_\perp}{\sqrt{\left(g_\parallel B_\parallel\right)^2+\left(g_\perp B_\perp\right)^2}}.
\label{eqRabiomgR}
\end{eqnarray}
In this expression, the two components of the magnetic field are given by $B_\parallel=B\cos\theta$ and $B_\perp=B\sin\theta$, 
where $\theta$ is the angle of the magnetic field measured with respect to the growth plane. 
By absolute value, $\omega_{\rm R}$ is largest at
\begin{eqnarray}
\theta &=& \pm\arctan\left(\sqrt{\left|\frac{g_\parallel}{g_\perp}\right|}\right),
\end{eqnarray}
attaining
\begin{equation}
\omega_{\rm R} =\pm \frac{\mu_B V_{\rm ac}}{2\hbar}
\left[
\frac{1}{g_\parallel}\left(\frac{\partial g_\parallel}{\partial V_{\rm tg}}\right)-
\frac{1}{g_\perp}\left(\frac{\partial g_\perp}{\partial V_{\rm tg}}\right)
\right]
\frac{g_\parallel g_\perp B}{\left|g_\parallel\right| +\left|g_\perp\right|}.
\end{equation}

\acknowledgement
We acknowledge financial support from 
the Nanosciences Foundation (Grenoble, France), 
the Commission for a Marie Curie Carrer Integration Grant, 
the Austrian Science Fund (FWF) for a Lise-Meitner Fellowship (M1435-N30), 
the DOE under Contract No. DE-FG02-08ER46482 (Yale), 
the European Starting Grant program, 
and 
the Agence Nationale de la Recherche. 
The authors thank J.W.G. van den Berg and S. Nadj-Perge for useful discussions.

\bibliography{bibliography_NA}

\begin{mcitethebibliography}{25}
\bibitem[lev()]{leverarm}
The lever arm of a gate electrode is the factor that relates a change of gate
  voltage to a corresponding shift of the subband energies.\relax
\mciteBstWouldAddEndPunctfalse
\mciteSetBstMidEndSepPunct{\mcitedefaultmidpunct}
{}{\mcitedefaultseppunct}\relax
\EndOfBibitem
\end{mcitethebibliography}
\end{document}